\documentstyle[11pt]{article}
\begin{document}
%
%
\begin{titlepage}
\vspace{28.5mm}
\begin{center}
{\huge\bf $\tau$ Decay and the \\ QCD infrared fixed point }
\vspace{29mm}\\
{\Large A. C. Mattingly }
\vspace{18mm}\\
{\large\it
T.W. Bonner Laboratory, Physics Department,\\
Rice University, Houston, TX 77251, USA}
\vspace{30mm}\\
{\bf Abstract:}
\end{center}

We apply the optimization procedure based on the Principle of Minimal
Sensitivity to the third-order calculation of $R_{\tau}$.
Since the effective couplant
remains finite, freezing to a value $\alpha_s/\pi = 0.26$ at low energies, we
can actually evaluate the defining integral of $R_{\tau}$ and compare the
optimized perturbation theory result to that of the optimized result obtained
after the integral has been evaluated using contour techniques.  The good
agreement shows that the optimization procedure is consistent and suggests
that the infrared fixed point is meaningful.
\end{titlepage}
\setcounter{page}{1}

\subsection{Introduction}
The inclusive semihadronic decay rate of the $\tau$ lepton, expressed as the
ratio
\begin{equation}
 R_{\tau} = \frac{\Gamma(\tau^{-} \rightarrow \nu_{\tau} + \mbox{hadrons})}
{\Gamma(\tau^- \rightarrow \nu_{\tau} e^- \bar{\nu}_e)},
\end{equation}
is a fundamental test of QCD.
The parton model gives a rough estimate by approximating hadronic decay as the
rate into quark-antiquark pairs:
$R_{\tau}  \approx N_c = 3$. Corrections to this approximation include both
perturbative and non-perturbative QCD, as well as electroweak corrections
\cite{braaten,pich}.
Here we will concentrate on the pertubative QCD corrections and compare
two different ways of applying the optimized perturbation theory to $R_{\tau}$.

Refs.\ \cite{braaten,pich} discuss in detail the theoretical calculation of
$R_{\tau}$, which has the following form:
\begin{equation}
 R_{\tau} =  \int^{M_{\tau}^2}_0 ds \; D(s) =
\int^{M_{\tau}^2}_0 ds \; \frac{2}{M_{\tau}^2}
\left(1 - \frac{s}{M_{\tau}^2} \right)^2 (1 + \frac{2s}{M_{\tau}^2})
\tilde{R}(s),
\label{integral}
\end{equation}
with
\begin{equation}
 \tilde{R}(s) = 3 (|V_{ud}|^2 + |V_{us}|^2) [1 + {\cal R}_{\tau}(s) ].
\label{rtild}
\end{equation}
Using the couplant $a \equiv \frac{\alpha_s}{\pi}$
the perturbative corrections
are expressed as a series, truncated to third-order:
\begin{equation}
 {\cal R}_{\tau} = a(1 + r_1 a + r_2 a^2).
\label{scr}
\end{equation}
{}From the recently re-done third-order calculation of Gorishny, {\it et al.}
\cite{new,r1calc} :
\begin{equation}
r_1(\overline{\rm MS},\mu=M_{\tau}) = 1.986 - 0.115 n_f, \hspace*{1cm}
\end{equation}
\begin{equation}
r_2(\overline{\rm MS},\mu=M_{\tau}) = -6.637 -1.200 n_f - 0.005 n_f^2,
\end{equation}
in the $\overline{\rm MS}$ scheme with the renormalization scale $\mu$ taken to
be $M_{\tau}$.
For three flavors, $r_1=1.64$ and $r_2=-10.28$.  [Note this differs
greatly from the
earlier result $r_2=93.98$ \cite{braaten} based on the
erroneous calculation of \cite{old}.]

Historically the integral (\ref{integral})
has not been evaluated directly because
it was expected that the QCD couplant would become large at small $s$, making a
perturbative prediction impossible.
Instead, $R_{\tau}$ was re-expressed as a contour integral in the complex $s$
plane with the contour running clockwise around the circle of
$|s|= {M_{\tau}^2}$,
thus avoiding the small-$s$ region.  To evaluate the integral, the coupling
constant is expanded in powers of $\alpha_s(M_{\tau})$ \cite{braaten,pich}.
The final result (for three flavors) is then, in the $\overline{\rm
MS}(\mu=M_{\tau})$ scheme, \cite{new}:
\begin{equation}
R_{\tau} = 3 (|V_{ud}|^2 + |V_{us}|^2) [1 + a + 5.20 a^2 + 26.37 a^3].
\label{rtau}
\end{equation}

\subsection{Optimized Perturbation Theory}

Based on the principle of minimal sensitivity,
optimized perturbation theory (OPT) \cite{opt} finds the
renormalization scheme (RS) in which the result is least
sensitive to changes in the RS
parameters.
For a detailed discussion of the OPT method see \cite{opt}.  The application
of Ref.\ \cite{us} to
the {\mbox{${R_{e^+e^-}}$}} ratio in third order is easily adapted to the
present case.
In effect, the difference between QCD corrections to the ratio
of the $e^+e^-$ hadronic cross section and the $\tau$ hadronic decay
involves taking $(\sum Q_f)^2 =0$ (in $r_2({\overline{MS}})$)
and replacing $3 \sum Q_f^2$ by  $3 \sum V_{ff}^2 \approx 3$.

As previously explained in \cite{kubo} the question of an infrared fixed point
can be addressed by solving the optimization equations and requires a
third-order calculation to determine the RS invariant $\rho_2$, where
\begin{equation}
\rho_2 \equiv r_2 + c_2 -(r_1+\frac{1}{2}c)^2,
\end{equation}
and $c$ and $c_2$ are $\beta$-function coefficients;
\begin{equation}
\mu \frac{\partial a}{\partial \mu} \equiv \beta(a) =
-b a^2 (1+ca+c_2 a^2+...).
\end{equation}
If $\rho_2$ is negative a positive fixed point, $\bar{a}^*$, exists,
and the more negative $\rho_2$ is, the smaller that $\bar{a}^{*}$ will be.
Just as in the {\mbox{${R_{e^+e^-}}$}} case, here $\rho_2$ is
negative and an infrared fixed point exists.
Our analysis shows that the numerical solutions to the optimization equations
tend towards that fixed-point solution.  Because of this ``freezing'' of the
coupling constant to a small finite value, the perturbative expansion of
${\cal R}_{\tau}(s)$ does not diverge and the integral (\ref{integral}) can be
evaluated numerically.  Thus there are two ways of applying optimized
perturbation theory to the problem of evaluating $R_{\tau}$:
(i) optimization of the integrated expression for $R_{\tau}$,
{\it i.e.} Eq.\ (\ref{rtau}),
and (ii)
optimization of
the perturbative expansion inside the integral down to low energies and then
integrating the result.  The second method gives a perturbative prediction for
the differential decay rate.

Method (i) involves applying
optimized perturbation theory to the $\overline{\rm MS}$
result for $R_{\tau}$, (\ref{rtau}), and was previously
done by Chyla {\it et al.}, \cite{Chyla}.
Here we take $V_{ud}=0.9747$ and $V_{us}=0.218$ from the particle data book
\cite{booklet}.
For three flavors and $s={M_{\tau}^2}$
[using $\Lambda^{(3)}_{\overline{MS}}=280$ MeV \cite{note}, and $M_{\tau}
=1.777$ GeV \cite{dallas}]
we get the following optimum values: $\bar{a}^{(3)}(M_{\tau}) = 0.164$,
$\bar{r}_1 = -0.49$, $\bar{r}_2 = 2.98$ and
$\bar{R}_{\tau}=3.48$.

Now we can compare this result to method (ii) where we have to solve the
optimization equations for ${\cal R}_{\tau}(s)$ in the range
${M_{\tau}^2} > s > 0$
and evaluate
the integrand of Eq.\  (\ref{integral}) explicitly.  Note
that the perturbative
expression to be optimized in this case is (\ref{scr}) which has different
coefficients than (\ref{rtau}).
At $s={M_{\tau}^2}$ the relevant number of flavors is
three (and again we will take
$\Lambda^{(3)}_{\overline{MS}}=280$ MeV for comparison to method (i)) but
at lower energies there are only two active flavors, so we must
`match' to a 2-flavor theory with an appropriate $\Lambda$.
We follow the method outlined in \cite{us}
by adjusting $\Lambda^{(2)}_{\overline{MS}}$ so that ${\cal R}_{\tau}$\ is
continuous at the matching point, which we take to
be $\sqrt{s}=2 m_s$,
(with a current quark mass of 199 MeV $\pm$ 33 \cite{booklet}).
This leads to a value of
$\Lambda^{(2)}_{\overline{MS}}=250$ MeV.
Another relevant point is that the term
$|V_{us}|^2$ must be set to zero
below the $s$,$u$ kinematic threshold at $\sqrt{s}=m_s + m_u$, since for
$\sqrt{s}$ below this value the virtual $W$ can decay only to an $u$,$d$ pair.

As with {\mbox{${R_{e^+e^-}}$}},
solving the optimization equations is straightforward except
for the slow convergence of the iteration method at low energies.
Numerical solutions were obtained down to
$\sqrt{s}=0.10$ GeV, where the result joins
smoothly to the analytic result for the $ \sqrt{s} \rightarrow 0$
limit \cite{kubo,us}.
In Fig.\ 1 we show the
optimized couplant $\bar{a}$ as a function of $\sqrt{s}$.  Also shown is the
optimized third-order correction $\bar{\cal R}_{\tau}(s)$.
At zero energy we have
the infrared-fixed point solution for two flavors.  In Table 1 we list the
infrared values for 2 and 3 flavors. [In the {\mbox{${R_{e^+e^-}}$}}
case, the results $\rho_2$
and $a^*$ are identical for 3 flavors (since $(\sum Q_f) =0 $ for $u$,$d$,$s$
quarks), but slightly different for 2 flavors ($\rho_2=-10.91, a^*=0.263$).]

{}From the optimized result for $R_{\tau}(s)$ we
can obtain $D(s)$ from Eqs.\ (\ref{integral}),(\ref{rtild}).
This is plotted in Fig.\ 2.  Integrating this from 0 to ${M_{\tau}^2}$ gives
$R_{\tau} = 3.44$, which agrees well with the result 3.48 obtained with
method (i).

$D(s)$ represents a perturbative prediction for the differential decay rate
of $d\Gamma(\tau^- \rightarrow \nu_{\tau} + \mbox{hadrons}(s))/ds$, normalized
by $\Gamma(\tau \rightarrow \nu_{\tau}e^-\bar{\nu_e})$, for $\tau$ decays to
hadrons with an invariant mass of $\sqrt{s}$ \ \cite{led}.
We do not expect this prediction to be right, of course.
The corresponding experimental
quantity would presumably show structure due to
meson thresholds and resonances, particularly $\pi$ and $a_1$.  However our
hypothesis is that the perturbative prediction is meaningful
in that if both the
data and the prediction are `smoothed' in some suitable fashion then they will
agree.  This hypothesis is supported by the results of Ref.\cite{us} which
finds excellent agreement when Poggio-Quinn-Weinberg (PQW) smearing is applied
to both the experimental data and the theoretical prediction for $R_{e^+e-}$.
We suggest that PQW smearing applied to $\tilde{R}(s)$, related to $D(s)$ by
(\ref{integral}), will also give good agreement in the $\tau$ case.
Unfortunately, it does not seem possible to test this prediction at present,
since existing $\tau$-decay data give inadequate information about neutral
hadrons.  It is a bold prediction because it implies that the vector and
axial-vector contributions, although dominated by different hadrons, must
become the same after smearing.  While the smearing is quite drastic (since it
must smooth out the peak structure in the data) the theory/experiment
comparison it allows is quantitative and highly non-trivial.

\subsection{Theoretical Uncertainties}
To answer the question of how good the agreement is between the two methods we
need to ask where the uncertainty lies in the calculation.  Other than
the choice of $\Lambda^{(3)}_{\overline{MS}}$ (which was the same for both
methods) there is only one uncertainty in method (i): the truncation of the
perturbative series.  Following the argument of \cite{us} we estimate the
error as $|\bar{r}_2 \bar{a}^3|$,
and thus for method (i):
\begin{equation}
R_{\tau}^{(i)} = 3.48 \pm 0.04.
\end{equation}

In method (ii) there are two sources of error: truncation of the
perturbation series and uncertainty in the input parameters (the strange quark
mass and $\Lambda^{(2)})$.  The series-truncation error was again estimated as
$|\bar{r}_2 \bar{a}^3|$ and then integrated as in Eq.\ (\ref{rtild}).
This had an uncertainty of about $\pm 0.10$.

As explained by Marciano \cite{Marciano} the $\Lambda_{\overline{ \rm MS}}$
of our effective theories (massless quarks with different numbers of flavors)
must be matched at the thresholds to correspond to a single, underlying ``full
QCD'' theory.  The exact procedure for doing this, though, is unknown; should
one make $\cal{R}$ continuous or $\alpha_s$ continuous across the threshold?
The differences between the two procedures are small, only about 5 MeV,
but we make the following estimate to gauge
the size of error: if $\Lambda^{(2)}_{\overline{MS}}$ increases to 260 MeV
then $R_{\tau}$ increases by only 0.002, which shows that the perturbative
corrections are insensitive to this uncertainty in
$\Lambda^{(2)}_{\overline{MS}}$.  This agrees
with the $e^+e^-$ analysis, \cite{us}, which found that below the $s$-quark
threshold {\mbox{${R_{e^+e^-}}$}} was most sensitive to the fixed-point
solution.

The estimated error on the current strange quark's mass is about 20\%.  If the
mass of the strange quark is decreased to 166 MeV, the 3 flavor region now runs
down to 332 MeV.  The effects of this on ${\cal R}_{\tau}$ are barely
discernable and do not significantly modify the integral.  The moving of the
threshold also affects the value of $\Lambda^{(2)}_{\overline{MS}}$.  For both
limits of the strange quark mass the $\Lambda^{(2)}_{\overline{MS}}$ increases
to about 260 MeV.  As we found above this also has little effect on the error.
The net effect of changing the strange quark mass is estimated to be $\pm
0.006$.
Therefore with method (ii) we estimate:
\begin{equation}
R_{\tau}^{(ii)} = 3.44 \pm 0.11.
\end{equation}
As one can see the two methods agree well within the error estimates
of method (ii).

\subsection{Comparison to Experiment}
$R_{\tau}$ can be found by measuring the leptonic branching fractions,
$B_e$, $B_{\mu}$ \cite{pich}
\begin{equation}
R_{\tau}^{exp,B} = 3.66 \pm 0.05.
\end{equation}
Independently $R_{\tau}$ can also be found by measuring the
total decay rate:
\[ R_{\tau}^{exp,\Gamma} \equiv \frac{\Gamma_{\tau} -
\Gamma_{\tau \rightarrow e} -\Gamma_{\tau \rightarrow \mu}}
{\Gamma_{\tau \rightarrow e}} \]
where $\Gamma_{\tau \rightarrow l} = \Gamma(\tau^{-} \rightarrow
\nu_{\tau}l^{-} \bar{\nu}_l) $ can be calculated theoretically with great
accuracy because it is a purely electroweak process. Using a mean life time
of $1/\Gamma = \tau_0 = (0.3025 \pm 0.0059) \times 10^{-12}$s
Ref.\ \cite{pich}
quotes $ R_{\tau}^{exp, \Gamma} = 3.32 \pm 0.12$.
Taking the average of these two values gives:
\begin{equation}
R_{\tau}^{exp} = 3.61 \pm 0.05.
\end{equation}

To compare to experiment we need to estimate the size of error due to the
choice of $\Lambda^{(3)}_{\overline{MS}}$. Increasing the value by 50 MeV to
330 MeV leads to a 2\% increase in method (i) and 1\% increase in method (ii).
In Table 2 we list the actual values for the case of increasing and decreasing
$\Lambda^{(3)}_{\overline{MS}}$ by 50 MeV.  Note that method (i) is more
sensitive to these changes than method (ii).

It is
generally accepted that non-pertubative corrections are small, decreasing $R$
by only 0.5\% $R_{\tau}^{pert}$ \cite{csli}.
Electroweak corrections are estimated to increase $R$ by 2\%.
Therefore we get the following estimates for the two methods:
\begin{equation}
R_{\tau}^{(i)} = 3.48 - 0.02 + 0.07 = 3.53 \pm 0.12,
\end{equation}
\begin{equation}
R_{\tau}^{(ii)} = 3.44 - 0.02 + 0.07 = 3.48 \pm 0.15.
\end{equation}
Both methods give very good results, which are near the experimental average
and well within the errors.

\subsection{Conclusion}
Using optimized perturbation theory we have shown that the infrared fixed point
and the `freezing' of the strong coupling constant at low energies can be used
to evaluate the $R_{\tau}$ integral directly.  Comparison to the usual OPT
method of optimizing the $\overline{\rm MS}$ result is excellent showing the
consistency of OPT.   Futhermore
both methods give good agreement to the experimental value.

\subsection*{Acknowledgements}
I would like to thank Paul Stevenson and Ian Duck for helpful comments.
This work was supported in part by the U.S. Department of Energy under Contract
No.\ DE-FG05-92ER40717.


\newpage
\begin{tabular}{cccc} \hline
$n_f$ & $\rho_2$ & $a^{*}$ & ${\cal R}_{\tau}(0)$ \\ \hline
2 & -10.83 & 0.264 & 0.332 \\
3 & -12.21 & 0.244 & 0.303 \\ \hline
\end{tabular}

\mbox{}
\vskip 1cm
{\bf Table 1:} Infrared fixed point values for 2 and 3 flavors.
\vskip 1in

\begin{tabular}{c|c|c} \hline
$\Lambda^{(3)}_{\overline{MS}}$ (MeV) & ${\cal R}_{\tau}^{(i)}$ &
${\cal R}_{\tau}^{(ii)}$ \\
& & \\ \hline
330 & 3.56 $\pm$ 0.06 & 3.48 $\pm$ 0.13 \\
280 & 3.48 $\pm$ 0.04 & 3.44 $\pm$ 0.10 \\
230 & 3.41 $\pm$ 0.02 & 3.40 $\pm$ 0.08 \\ \hline
\end{tabular}

\mbox{}
\vskip 1cm
{\bf Table 2:} Comparison of methods (i) and (ii) for different
$\Lambda^{(3)}_{\overline{MS}}$.  Errors are estimated by $|r_2 a^3|$,
integrated for method (ii).

\newpage
\noindent{\bf Fig.\ 1.} The optimized third-order results for
$\bar{a} = \alpha_s/\pi$
and $\bar{{\cal R}}_{\tau}^{(3)}(s)$.
The vertical line indicates where the $N_f=2$ and $N_f=3$ effective theories
are matched.  Error bars for
$\bar{\cal R}_{\tau}^{(3)}(s)$ are $\pm |\bar{r}_2 \bar{a}^3|$.

\vskip 1in

\noindent{\bf Fig.\ 2.} Integrand of $R_{\tau}$; {\it i.e.} the differential
decay rate $d\Gamma(\tau^- \rightarrow \nu_{\tau} + \mbox{hadrons}(s))/ds$
for decays into hadrons with invariant mass-squared $s$, divided by
$\Gamma(\tau^- \rightarrow \nu_{\tau} e^- \bar{\nu_e})$. The `jump' at
$\sqrt{s} \approx 200$ MeV occurs because the term $|V_{us}|^2=0$ below the
$s,u$ kinematic threshold.


\begin{thebibliography} {99}
\bibitem{note}
This value was used by \cite{Chyla} and agrees well with the 285 MeV found by
requiring {\mbox{${R_{e^+e^-}}$}}
be continuous across thresholds and using the world average
value of $\Lambda^{(4)}_{\overline{MS}}=230$ MeV.  For further details see
\cite{us}.
\end{thebibliography}
\end{document}